\documentclass[showpacs,prl,onecolumn,aps,superscriptaddress,preprintnumbers,nofootinbib]{revtex4}
\usepackage[T1]{fontenc}
\usepackage[latin9]{inputenc}
\usepackage{amsmath,amssymb,bigints}
\usepackage{footmisc}
\usepackage{epsfig}
\usepackage{graphicx}
\usepackage{amsmath}
\usepackage{amsfonts}
\def\slashchar#1{\setbox0=\hbox{$#1$}     		
   \dimen0=\wd0                                 	
   \setbox1=\hbox{/} \dimen1=\wd1               	
   \ifdim\dimen0>\dimen1                        	
      \rlap{\hbox to \dimen0{\hfil/\hfil}}      	
      #1                                        	
   \else                                        	
      \rlap{\hbox to \dimen1{\hfil$#1$\hfil}}   	
      /                                         	
   \fi}

\renewcommand{\vec}{\boldsymbol}
\newcommand{\beq}{\begin{equation}}
\newcommand{\eeq}{\end{equation}}
\newcommand{\bea}{\begin{eqnarray}}
\newcommand{\eea}{\end{eqnarray}}
\newcommand{\baa}{\begin{array}}
\newcommand{\eaa}{\end{array}}

\def\eq#1{{Eq.~(\ref{#1})}}
\def\fig#1{{Fig.~\ref{#1}}}

\newcommand{\nn}{\nonumber}

\newcommand{\h}{\frac{1}{2}}

\newcommand{\Lb}{\left(}
\newcommand{\Rb}{\right)}
\def\pom{{I\!\!P}}

\renewcommand{\vec}[1]{\boldsymbol{#1}}

\def\pom{{I\!\!P}}

\begin{document}
\title{ Double parton interaction: the values of $\sigma_{\rm eff}$}
\author{E. ~Gotsman}
\email{gotsman@post.tau.ac.il}
\affiliation{Department of Particle Physics, School of Physics and Astronomy,
Raymond and Beverly Sackler
 Faculty of Exact Science, Tel Aviv University, Tel Aviv, 69978, Israel}
 \author{ E.~ Levin}
 \email{leving@tauex.tau.ac.il, eugeny.levin@usm.cl}
  \affiliation{Department of Particle Physics, School of Physics and Astronomy,
Raymond and Beverly Sackler
 Faculty of Exact Science, Tel Aviv University, Tel Aviv, 69978, Israel} 
 \affiliation{Departemento de F\'isica, Universidad T\'ecnica Federico Santa Mar\'ia, and Centro Cient\'ifico-\\
Tecnol\'ogico de Valpara\'iso, Avda. Espana 1680, Casilla 110-V, Valpara\'iso, Chile}

\date{\today}


\begin{abstract}
In this letter we  show that the two parton showers mechanism for
 $J/\Psi$ production, that has been discussed in Ref.\cite{LESI},
 leads to small values of $\sigma_{\rm eff}$ 
for the production of a pair of $J/\Psi$. We develop a simple two
 channel approach to  estimate  the values of $\sigma_{\rm eff}$,
 which produces   values   that are  in accord with the experimental 
data.

 \end{abstract}

\preprint{TAUP - 3038/19}

\maketitle

\subsection{Introduction}
\begin{figure}[ht]
    \centering
  \leavevmode
      \includegraphics[width=10cm]{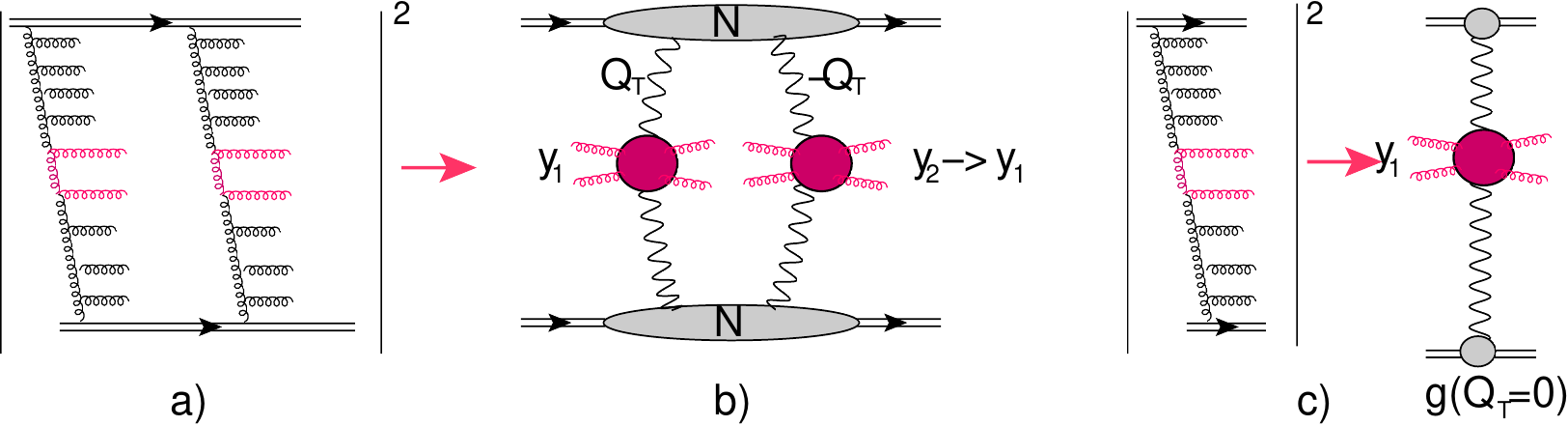}  
    \caption{Two parton showers contribution to two pairs of  back-to-back 
gluon
 jet production in hadron-hadron collisions (\fig{dpi4j}-a). Helical 
lines
 denote the gluons. The red helical lines  depict the jets. The wavy 
lines
 describe the BFKL Pomeron. \fig{dpi4j}-b  shows the Mueller
 diagram\cite{MUDI} for the double inclusive cross section, while
 \fig{dpi4j}-c is the Mueller diagram for the inclusive cross section.}
\label{dpi4j}
  \end{figure}

\begin{figure}[h]
    \centering
  \leavevmode
      \includegraphics[width=12cm]{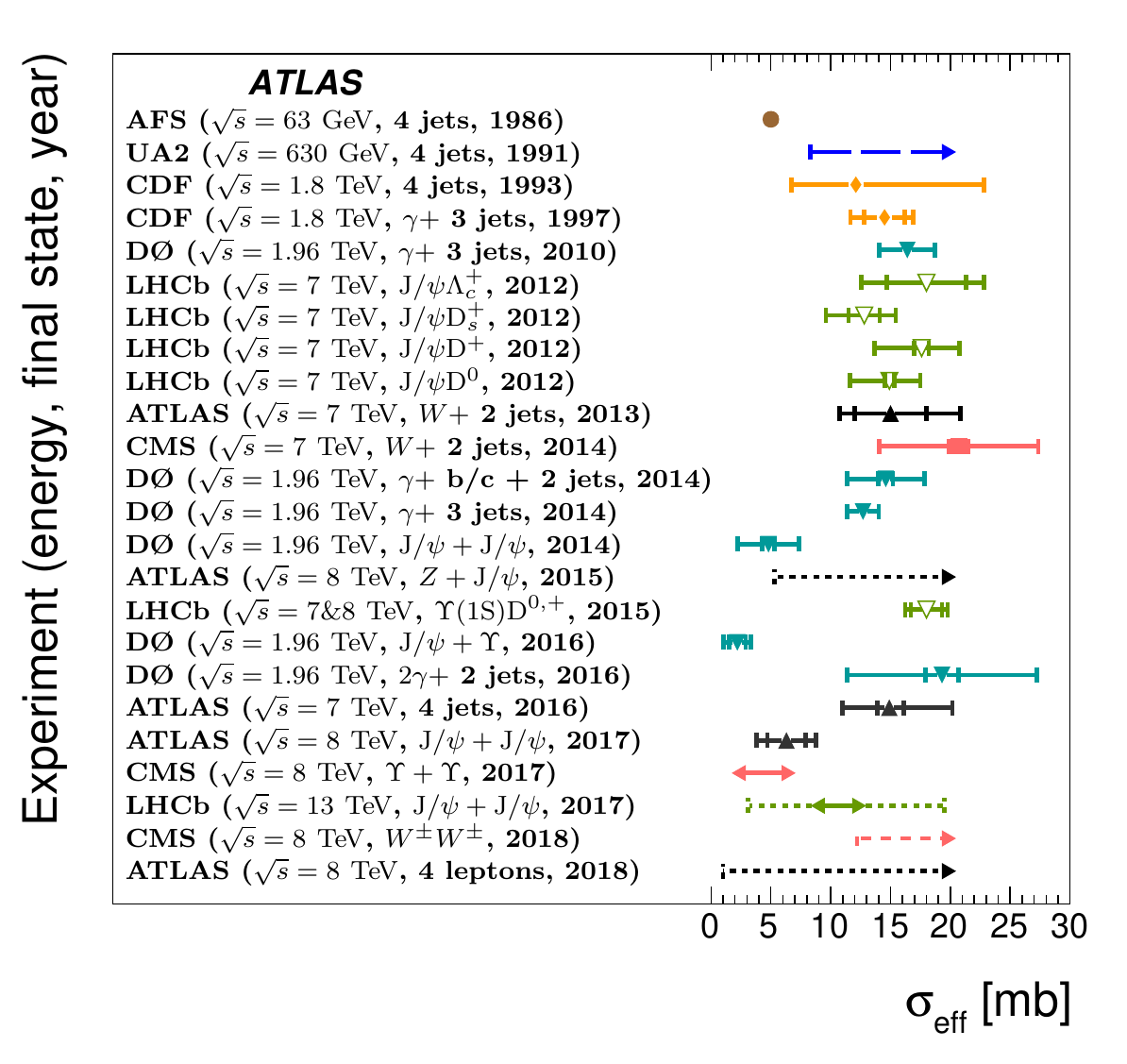}  
    \caption{Summary of measurements and limits on the effective cross
 section, determined in different experiments\cite{AFSC,UA2,CDF,CDF2,D0,LHCB,
ATLAS,CMS,ATLAS2,D01,LHCB1,ATLAS3,CMS1,ATLAS4,D03,ATLAS5,LHCB3,D04,CMS2,
ATLAS7,LHCB4,D05,D06,ATLAS6,CMS3} . 
   The measurements that were made by different experiments are denoted
 by different
symbols and colours. The figure is taken from Ref.\cite{ATLAS7}.}
\label{sigeff}
  \end{figure}

The double parton interaction has been under close  scrutiny  over 
the past
  three decades both by theoreticians (see Ref.\cite{DIGA} and references
 therein) and by the experimentalists\cite{AFSC,UA2,CDF,CDF2,D0,LHCB,ATLAS,CMS,ATLAS2,D01,LHCB1,ATLAS3,CMS1,ATLAS4,D03,ATLAS5,LHCB3,D04,CMS2,ATLAS7,LHCB4,D05,D06,ATLAS6,CMS3}.  The  fairly large cross sections for the 
double
 parton interaction at high energy  support the assumption, that a
 dense system of partons is produced in the proton-proton collisions
 at high energy. 
Such dense systems  of partons appears naturally  in the CGC/saturation
 approach to high energy QCD\cite{KOLEB},
which  permits  us to  consider  hadron-hadron, hadron-nucleus 
and
 nucleus-nucleus interactions from a unique point of view.
  As a quantitive measure of the strong double parton interaction,
 the value of $\sigma_{\rm eff}$ is used,  this was introduced by  
considering 
 the double inclusive cross sections of two pairs of back-to-back
 jets with momenta $p_{T,1}$ and $p_{T,2}$,   measured with
 rapidities  of two pairs ($y_1$ and $y_2$),  which are close to
 each other ($y_1 \approx y_2$, see \fig{dpi4j}). These pairs
 can  only be produced  from two different parton showers.  The
 data were parameterized in the form
\beq \label{XSEFF}
\frac{d \sigma}{d y_1 d^2 p_{T,1} d y_2 d^2 p_{T,2}}
 \,\,=\,\,\frac{m}{2 \sigma_{eff}}\,\frac{d \sigma}{d
 y_1 d^2 p_{T,1}}\frac{d \sigma}{d y_2 d^2 p_{T,2}} 
\eeq
where $m =2 $ for different pairs of jets and $m=1$ for
 identical pairs. The values of $\sigma_{\rm eff}$ are
  shown in \fig{sigeff} for different final states.

One can see from this figure, that in spite of  large errors,
 pairs $J/\Psi +J/\Psi $ and $
J/\Psi  + \Upsilon$  have  a small value of $\sigma_{\rm eff} $
 of about 5 mb, while other final states lead to $\sigma_{\rm eff} \, of
 \approx $ 15 mb.

In this letter we wish to show that the two parton shower mechanism
 for $J/\Psi$ production, that has been discussed in Ref.\cite{LESI},
 leads to small values of $\sigma_{\rm eff}$, and also to develop a simple
 two channel approach to  estimate  the values of $\sigma_{\rm 
eff}$.

\begin{boldmath}
\subsection{$\sigma_{\rm eff}$ in the BFKL Pomeron calculus}
\end{boldmath}
The double inclusive cross section for two pairs of the jets
 shown in \fig{dpi4j} can be written in the form (see \fig{dpi4j}-b) 
\beq\label{DI1}
\frac{d \sigma}{d y_1 d^2 p_{T,1} d y_2 d^2 p_{T,2}} \,\,=\,\,\int\frac{ d^2\,Q_T}{4 \pi^2} \,\frac{ N^2\Lb Q_T\Rb}{g^4\Lb Q_T=0\Rb}\frac{d \sigma}{d y_1 d^2 p_{T,1}}\Lb Q_T\Rb \frac{d \sigma}{d y_2 d^2 p_{T,2}}\Lb Q_T\Rb\eeq

In \eq{DI1} $ \frac{d \sigma}{d y_1 d^2 p_{T,1}}\Lb Q_T\Rb$ 
 describes the production of two  back-to-back jets from the BFKL
 Pomeron\cite{BFKL} and can be written in the form:
\beq \label{DI10}
\frac{d \sigma}{d y_1 d^2 p_{T,1}}\Lb Q_T\Rb\,\,=\,\,G^{\rm BFKL}_\pom\Lb
 Y - y_1,k_T,Q_T\Rb \bigotimes \sigma_{\rm hard} \bigotimes
 G^{\rm BFKL}_\pom\Lb  y_2, \vec{p}_{1,T} + \vec{p}_{2,T} -
 \vec{k}_T; Q_T\Rb\eeq
where $\bigotimes$ stands for all necessary integrations and   $  Q_T  $
denotes the transverse momentum. 
 The amplitude
of the BFKL Pomeron $G^{\rm BFKL}_\pom\Lb Y - y_1,k_T,Q_T\Rb$
can be simplified if we take into account that the $\vec{Q}_{T}$
dependence of the BFKL Pomeron is determined by the size of the largest
of the interacting dipoles~\footnote{The fact that the $Q_{T}$ dependence
 is determined by the size of
the largest dipole stem from the general features of the BFKL Pomeron.
Indeed, the eigenfunction of the BFKL Pomeron in  coordinate space
is equal to \cite{LIP} 
\begin{equation}
N\left(r,r';b\right)\,\,=\,\,\left(\frac{r^{2}\,\,r'^{2}}{\left(\vec{b}-\frac{1}{2}\left(\vec{r}-\vec{r}'\right)\right)^{2}\,\left(\vec{b}+\frac{1}{2}\left(\vec{r}-\vec{r}'\right)\right)^{2}}\right)^{\gamma}\label{BFKLBDEP}
\end{equation}
where $b$ is the conjugate variable to $Q_{T}$. From Eq.~(\ref{BFKLBDEP})
one can see that the typical value of $b$ is of the order of the
larger of $r$ and $r'$. In our process $r'$ is of the order of
$R_{h}$, where $R_{h}$ denotes the radius of the nucleon. The value
of $1/r$ is of the order of the mass of the heavy quark $m_{c}$,
or the saturation scale $Q_{s}$ and, therefore, turns out to be much
larger than $1/R_{h}$, and can be neglected.  Note, that the typical
 values of $k_T \propto 1/r$\label{fnt}}.
 Indeed, in this case  $Q_{T}\approx1/R_{h}\,\,\ll\,\,k_T (p_T)$ and
  the $Q_T$ dependence  can be described 
 by $g\left(Q_{T}\right)$ in \eq{FD3}, which
has a non-perturbative origin and, in practice, has to be taken
from the experiment.
\begin{equation} 
G_\pom^{\rm BFKL}\Lb Y - y_1,k_T,Q_T\Rb\,\,\approx\,\,G_\pom^{\rm BFKL}\Lb Y - y_1,k_T,Q_T=0\Rb\,g\Lb Q_T\Rb\label{FD3}
\end{equation}
 \eq{FD3} has a simple interpretation in the framework of the BFKL
 Pomeron calculus (see Ref.\cite{KOLEB} for a review): $G_\pom^{\rm BFKL}$
 describes the BFKL Pomeron Green's function, while $g\Lb Q_T\Rb$ denotes 
the
 vertex of interaction of the Pomeron with the hadron.

\begin{figure}[ht]
    \centering
  \leavevmode
      \includegraphics[width=14cm]{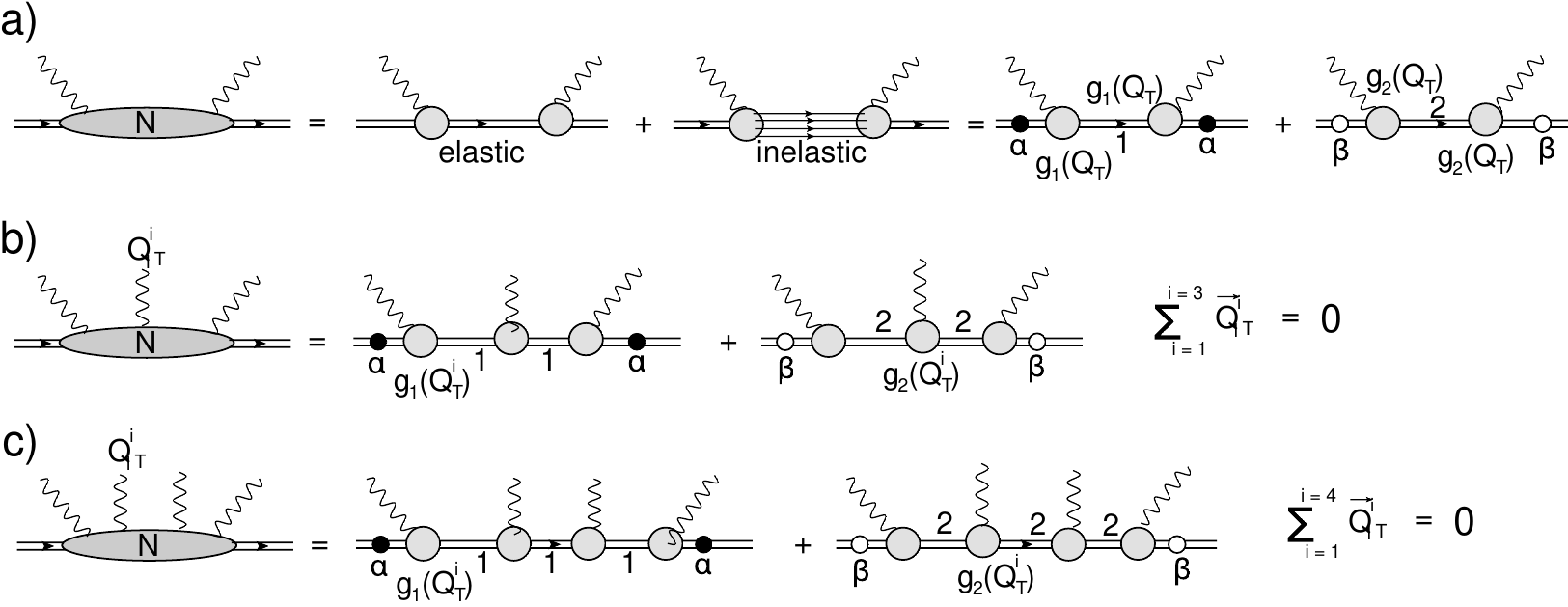}  
    \caption{The structure of the amplitude $N$ in the two channel model.}
\label{n}
  \end{figure}

For the  exchange of two BFKL Pomerons in \eq{DI1} we assume that dependence
 on $Q_T \sim 1/R_h$ is described by the function $N\Lb Q_T\Rb$, which can 
be
 treated as a phenomenological amplitude of  the  interaction of two 
BFKL Pomerons
 with the hadron. It is worth mentioning that the first contribution to $N\Lb
 Q_T\Rb = g^2\Lb Q_T\Rb$, which corresponds to the contribution  of 
the
  eikonal rescattering on the hadron to this amplitude  (see 
\fig{n}-a1).
 Therefore,  we  have

\bea \label{DI12}
\frac{d \sigma}{d y_1 d^2 p_{T,1} d y_2 d^2 p_{T,2}} &=&\int\frac{ d^2\,Q_T}{4 \pi^2} \, N^2\Lb Q_T\Rb\times\,\Bigg\{G^{\rm BFKL}_\pom\Lb  Y - y_1,k_T, 0\Rb   \bigotimes \sigma_{\rm hard} \bigotimes
\,G^{\rm BFKL}_\pom\Lb  y_2, \vec{p}_{1,T} + \vec{p}_{2,T} -
 \vec{k}_T; 0\Rb \nn\\
 &\times&\, G^{\rm BFKL}_\pom\Lb  Y - y_1,k_T, 0\Rb  \bigotimes \sigma_{\rm hard} \bigotimes
  \,G^{\rm BFKL}_\pom\Lb  y_2, \vec{p}_{1,T} + \vec{p}_{2,T} -
 \vec{k}_T; \vec{Q}, 0\Rb \Bigg\}
   \eea  

Finally, with these assumptions  we can re-write \eq{DI1} as follows:

\beq \label{DI101}
\frac{d \sigma}{d y_1 d^2 p_{T,1} d y_2 d^2 p_{T,2}} \,\,=\,\,\int\frac{ d^2\,Q_T}{4 \pi^2} \,\frac{ N^2\Lb Q_T\Rb}{g^4\Lb Q_T=0\Rb}\frac{d \sigma}{d y_1 d^2 p_{T,1}}\Lb Q_T=0\Rb \frac{d \sigma}{d y_2 d^2 p_{T,2}}\Lb Q_T=0\Rb
\eeq

 \eq{DI10} takes the form:
\beq \label{DI102}
\frac{d \sigma}{d y_1 d^2 p_{T,1}}\Lb Q_T=0\Rb\,\,=\,g^2\Lb Q_T=0\Rb
\,\Bigg\{G^{\rm BFKL}_\pom\Lb
 Y - y_1,k_T,0\Rb \bigotimes \sigma_{\rm hard} \bigotimes
 G^{\rm BFKL}_\pom\Lb  y_2, \vec{p}_{1,T} + \vec{p}_{2,T} -
 \vec{k}_T; 0\Rb\Bigg\}\eeq
which leads to

\beq \label{DI11}
\frac{1}{ \sigma_{\rm eff}} \,=\,\int \frac{d^2\,Q_T}{4 \pi^2} \,\frac{ N^2\Lb Q_T\Rb}{g^4\Lb Q_T=0\Rb}
\eeq

  Note, that the factors, that stem from short distances and are
 denoted by $\{\dots\}$ in \eq{DI12} and \eq{DI102},
   cancel in \eq{DI11}.
   
 The principal feature of \eq{DI11} is that the value of
 $\sigma_{\rm eff}$ does not depend on the nature of the hard process.
 This  could be the di-jet production, the inclusive production of 
$J/\Psi$
 or any of  the  hard processes shown in Fig.2,  however  
\eq{DI11} 
should be
 the same. As one can see from Fig.2, that experimentally for vast
 number of different hard processes, we have more or less the same values
 of $\sigma_{\rm eff}$ .   However, for $J/\Psi$ production  the
 value of $ \sigma_{\rm eff}$  turns out to be smaller than for other
 processes, in spite  of the fact that at first sight we can use the
 same \eq{DI11} for  its value. Actually,  $\sigma_{\rm eff} $
 characterizes  typical long distances that contribute to the
 hard processes. In the case of  single inclusive production we
 are able  to include these long distances in the experimental parton 
densities. However, for the double inclusive production we have to
 introduce more information  regarding these contributions. In our 
approach
 we introduce   the   functions $N\Lb Q_T\Rb$.

$N\Lb Q_T\Rb$  denotes the scattering amplitude of the BFKL Pomeron 
with the
 hadron integrated over all  produced mass, which has a non-perturbative
 origin, and should be taken from non-perturbative QCD, or from high 
energy
 phenomenology,  bearing in mind the current embryonic stage of 
 the non-perturbative
 approach. \eq{DI1}  is   discussed in more detail in Ref.\cite{LERE}.
The amplitude $N(Q_T)$ has  a complex structure and can be viewed as
 the sum of the elastic contribution and the inelastic one, and should
 be summed over  the  entire range of the produced mass (see 
\fig{n}-a).


In the two channel approximation we replace the rich structure of the
 produced  states, by a single  state with the wave
 function $\psi_D$.
  The observed physical 
hadronic and diffractive states are written in the form 
\beq \label{MF1}
\psi_h\,=\,\alpha\,\Psi_1+\beta\,\Psi_2\,;\,\,\,\,\,\,\,\,\,\,
\psi_D\,=\,-\beta\,\Psi_1+\alpha \,\Psi_2;~~~~~~~~~
\mbox{where}~~~~~~~ \alpha^2+\beta^2\,=\,1;
\eeq 

Functions $\psi_1$ and $\psi_2$  form a  
complete set of orthogonal
functions $\{ \psi_i \}$ which diagonalize the
interaction matrix ${\bf T}$
\beq \label{GT1}
A^{i'k'}_{i,k}=<\psi_i\,\psi_k|\mathbf{T}|\psi_{i'}\,\psi_{k'}>=
A_{i,k}\,\delta_{i,i'}\,\delta_{k,k'}.
\eeq

Bearing \eq{GT1} in mind, we can write the following expression for
 the amplitude $N$ (see \fig{n}-a):
\beq \label{N}
N\Lb Q_T\Rb\,\,=\,\,\alpha^2 g^2_1\Lb Q_T\Rb\,\,+\,\,\beta^2 g^2_2\Lb Q_T\Rb
\eeq
    In \eq{N} the first term corresponds to  the contribution of 
the 
state 
$\Psi_1$ and the second of the state $\Psi_2$, to the amplitude
 of the interaction of two Pomerons with the nucleon.

     \begin{figure}[ht]
    \centering
  \leavevmode
      \includegraphics[width=14cm]{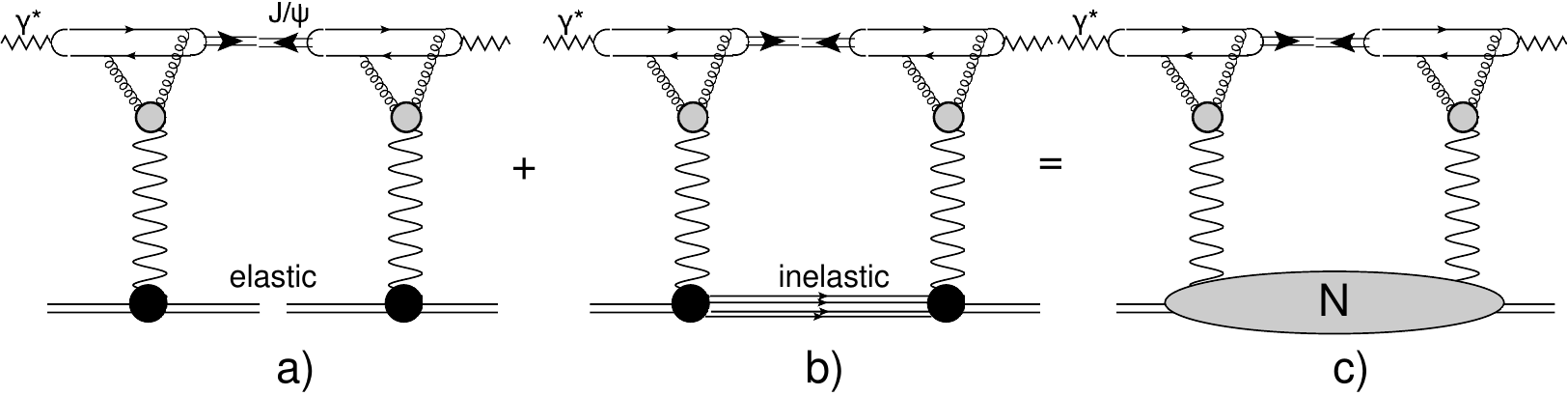}  
    \caption{Diffractive production of  the $J/\Psi$ meson in deep 
inelastic
 scattering at low $x$. Wavy lines denote  the BFKL Pomerons.}
\label{nd}
  \end{figure}

 $g_i(Q_T)$ are phenomenological functions, while $\alpha$ is a
  constant whose value we  determine by  fitting to the experimental
 data. Fortunately, we can determine the values of $g_i(Q_T)$ and
   $\alpha$      by considering the diffractive
 production of $J/\Psi$ in DIS\cite{H1DD,H1DD1,ZEUS1}. 
These data indicate that the diffractive production of $J/\Psi$ has
 different slopes in $Q_T$,  for elastic (see \fig{nd}-a) and
 for inelastic  (see \fig{nd}-b) production. For the reaction 
$\gamma^* + p \to J/\Psi + p$, $d \sigma_{el.\,diff}/d Q^2_T
 \propto \exp\Lb- B_{el}Q^2_T\Rb $ with $B_{el} \approx 5 \,GeV^{-2}$,
 while for $\gamma^* + p \to J/\Psi + X$ 
the slope turns out to be much smaller $d \sigma_{inel.\,diff}/d Q^2_T
 \propto \exp\Lb- B_{in} Q^2_T\Rb$ with $B_{in} \approx 1.69 \,GeV^{-2}$.
 The second  conclusion  from the  data is that the cross sections of 
elastic and
 inelastic $J/\Psi$ diffractive production are the same.
We can implement the described features of the diffractive production of
 $J/\Psi$  by introducing $ g_1\Lb Q_T\Rb = g_1 e^{ - \h B_1 \,Q^2_T}$ and
 $ g_2\Lb Q_T\Rb = g_1 e^{ - \h B_2 \,Q^2_T}$ and imposing the following
 restrictions on the parameters 
\beq \label{RES}
B_1 = B_{el}; ~~B_2=B_{in};~~~~\alpha^2 \frac{g^2_1}{B_1}\,=\,\beta^2 \frac{g_2^2}{B_2}
\eeq

 These equations for $B_i$ follows from the fact that  function
 $\Psi_1$ gives the main contribution to the for elastic diffractive
 production, while $\Psi_2$ is introduced to describe the inelastic 
 diffraction.

Plugging \eq{n} into \eq{DI1} we have
\beq \label{DI2}
\frac{1}{ \sigma_{\rm eff}} \,=\,\int \frac{d^2\,Q_T}{4 \pi^2}
 \,\frac{ \Lb \alpha^2 g^2_1\Lb Q_T\Rb  + \beta \,g^2_2\Lb Q_T\Rb \Rb^2}{\Lb
 \alpha^2 g_1\Lb 0\Rb + \beta^2 g_2\Lb 0\Rb\Rb^4 }
\eeq

\begin{figure}[ht]
    \centering
  \leavevmode
  \begin{tabular}{c c}
      \includegraphics[width=8.5cm]{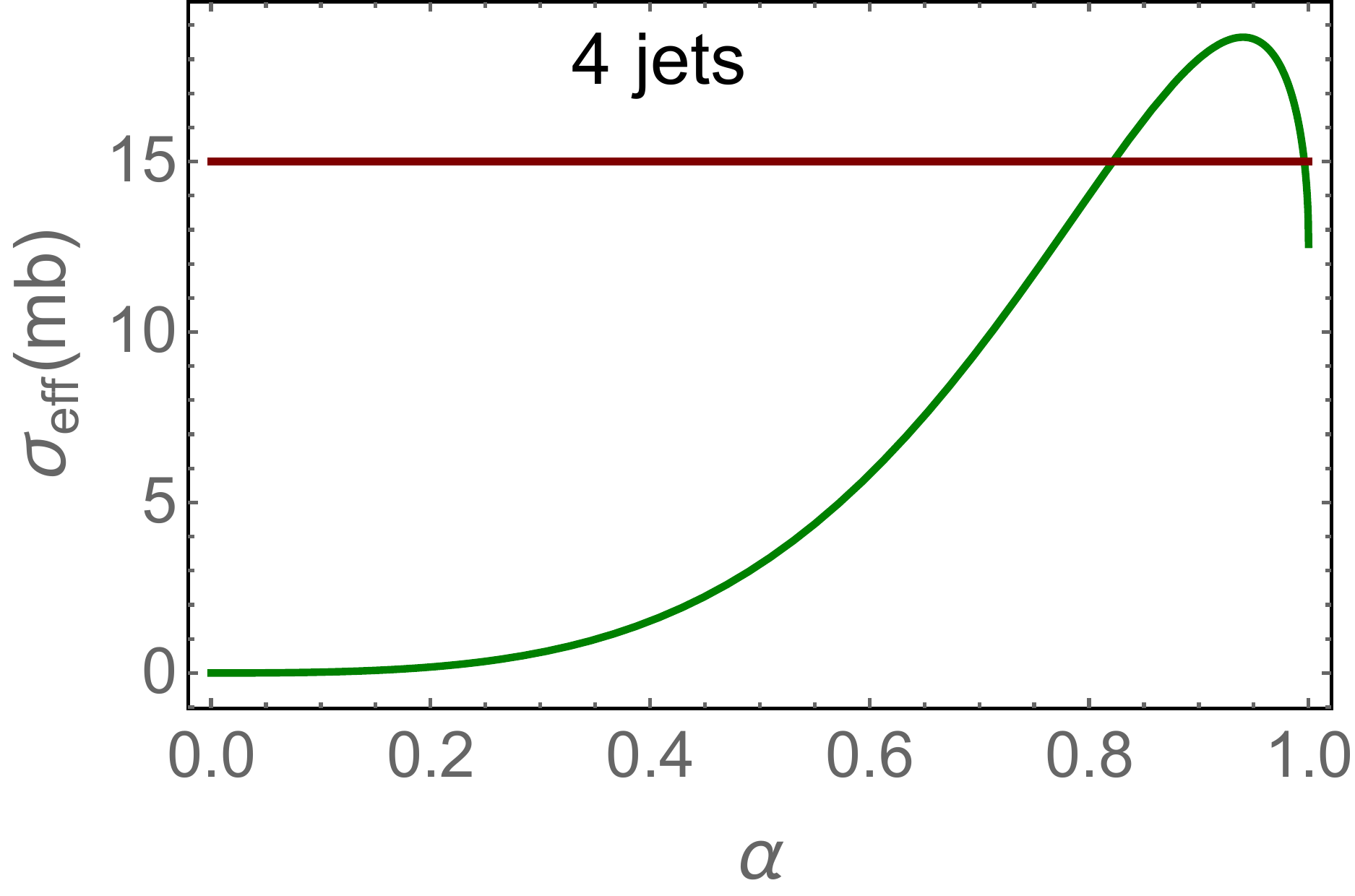} & 
  \includegraphics[width=8.5cm]{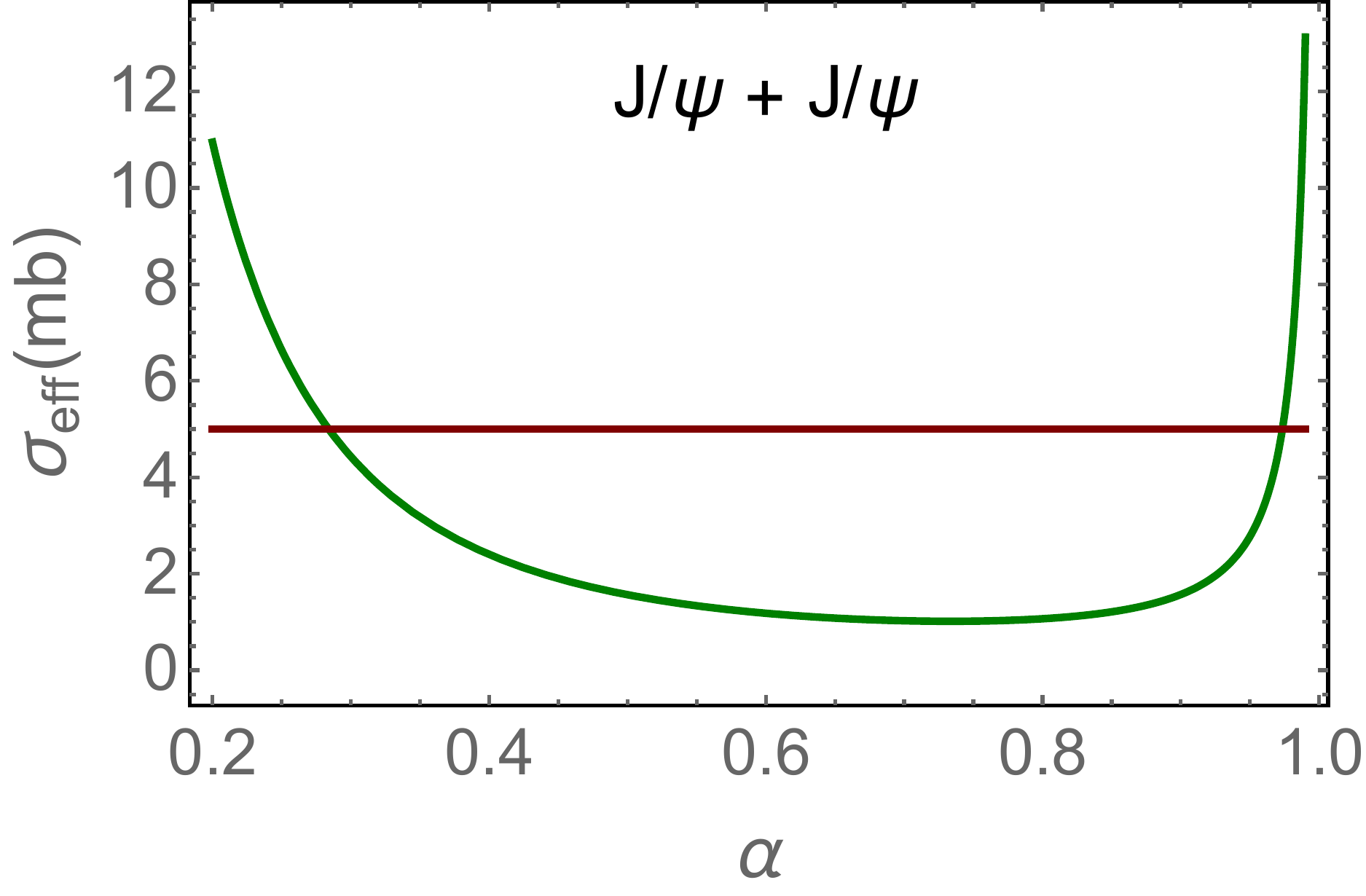}\\
      \fig{xs4j}-a &   \fig{xs4j}-b\\
      \end{tabular}
\caption{$\sigma_{\rm eff} $ for a  pair of back-to-back
 jets (see \fig{xs4j}-a and  \eq{DI1}),    and for pair of
 $J/\Psi$ (\fig{xs4j}-b). The red lines show
 $\sigma_{\rm eff}$ = 15 mb in \fig{xs4j}-a and
 $\sigma_{\rm eff}$ = 5 mb in \fig{xs4j}-b. }
\label{xs4j}
  \end{figure}

In \fig{xs4j} we plot the value of $\sigma_{\rm eff}$ for a pair of
 back-to-back jets.

 We reach the average experimental value
 of $\sigma_{\rm eff}=15 \,mb$ at $\alpha=0.8$. It should
 be noted that the information that we obtain from the
 diffractive production of $J/\Psi$, is enough to claim
 that $\sigma_{\rm eff} \leq 20\, mb$ in a two channel model.

~

\begin{boldmath}
\subsection{$\sigma_{\rm eff}$ for the production $J/\Psi$ and
 $\Upsilon$ pairs}
\end{boldmath}

\begin{boldmath}
\subsubsection{Two parton showers mechanism for $J/\Psi$ production}
\end{boldmath}

\begin{figure}[ht]
    \centering
  \leavevmode
      \includegraphics[width=10cm]{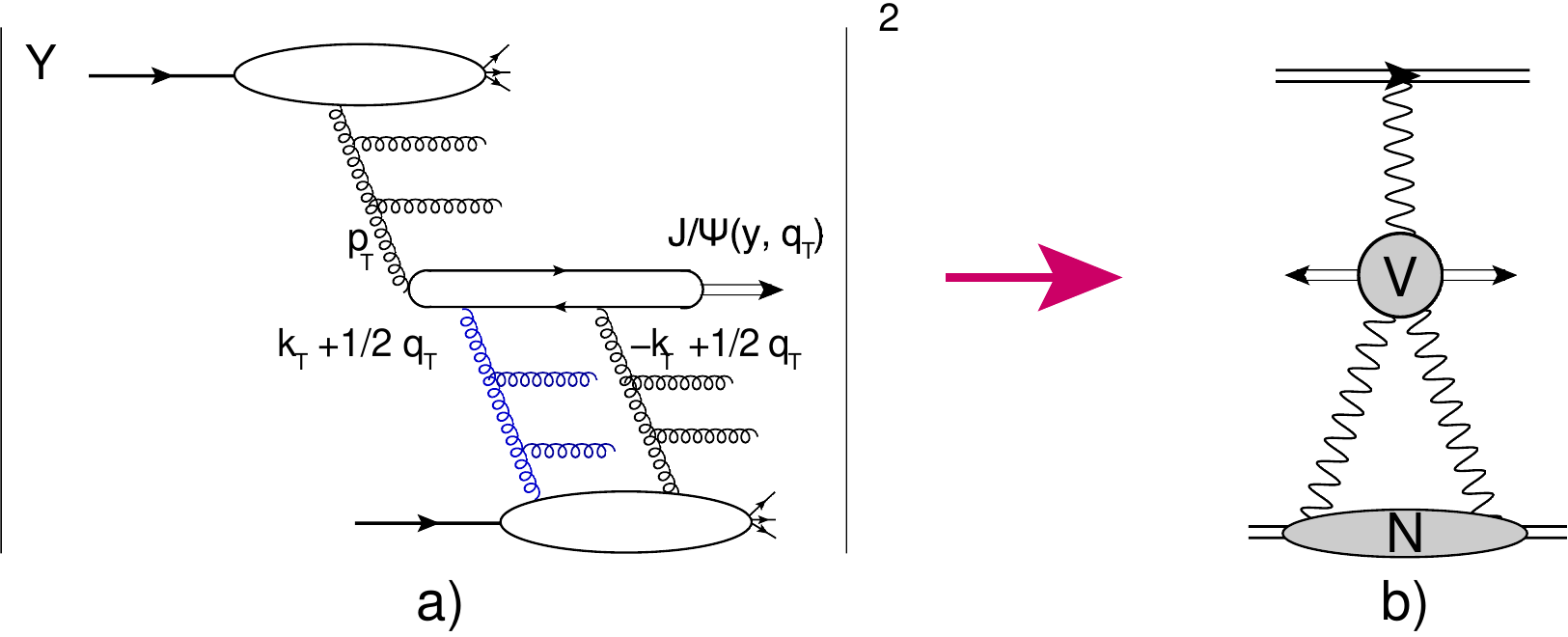}  
    \caption{The two parton shower mechanism for $J/\Psi$ production
 (see \fig{psi2sh}-a). \fig{psi2sh}-b shows the Mueller diagram\cite{MUDI}
 for the inclusive $J/\Psi$ production. The wavy lines describe
 the BFKL Pomerons. The vertex V is defined in \eq{FD51}.}
\label{psi2sh}
  \end{figure}

 In Ref.\cite{LESI} it is shown that the two parton shower mechanism
 of $J/\Psi$ production, which is  illustrated  in \fig{psi2sh}, 
gives the 
 description of  total and differential cross sections. In this paper
 we would like to show that this mechanism leads to a value of
 $\sigma_{\rm eff}$ which is much smaller than  the one of the
 previous section.  The formula for the single inclusive cross
 section of $J/\Psi$ prodiuction with transverse momentum $q_T$
 is  given in Ref.\cite{LESI} and has the following form
\begin{subequations}
\begin{eqnarray}
\frac{d\sigma\left(Y,\,Q^{2}\right)}{dy\,d^{2}q_{T}}\,\, & = & \,\,\frac{4\,C_{F}^{3}\,\bar{\alpha}_{S}^{3}}{(2\pi)^{4}}\,\int\frac{d^{2}Q_{T}}{(2\pi)^{2}}\,g\Lb 0 \Rb\,N\left(Q_{T}\right)\,\int d^{2}k_{T}d^{2}p_{T}\,\,G_{I\!\!P}^{\rm BFKL}\left(Y-y,\,p_{T},\,0\right)\,\label{FD5}\\
 & \times & I^{2}\left(\boldsymbol{k}_{T},\,\boldsymbol{q}_{T}\right)\,G_{{I\!\!P}}^{{\rm BFKL}}\left(y;\,\boldsymbol{k}_{T}+\frac{1}{2}\boldsymbol{q}_{T},\,0\right)\,G_{{I\!\!P}}^{{\rm BFKL}}\left(y,-\boldsymbol{k}_{T}+\frac{1}{2}\boldsymbol{q}_{T},\,0\right)\nonumber \\
 &=& \int d^2 Q_T\, g\Lb 0 \Rb\,N\left(Q_{T}\right)\,\Bigg\{ \int d^{2}k_{T}d^{2}p_{T}\,\,G_{I\!\!P}^{\rm BFKL}\left(Y-y,\,p_{T},\,0\right)\,\label{FD51}\\
&\times&V\Lb k_T,q_T\Rb\,G_{{I\!\!P}}^{{\rm BFKL}}\left(y;\,\boldsymbol{k}_{T}+\frac{1}{2}\boldsymbol{q}_{T},\,0\right)\,G_{{I\!\!P}}^{{\rm BFKL}}\left(y,-\boldsymbol{k}_{T}+\frac{1}{2}\boldsymbol{q}_{T},\,0\right)\Bigg\} \nn\\
&\propto& \,\,\int d^2 Q_T\, g\Lb 0 \Rb\,N\left(Q_{T}\right) \label{FD52}
 \end{eqnarray}
 \end{subequations}
 
 \eq{FD51} is rewritten in  \eq{FD5} in the form, which correspond 
the Mueller 
diagram of \fig{psi2sh}, and introduces the triple Pomeron vertex with
 emitted $J/\Psi$  ($V$ in \fig{psi2sh}) in the explicit form. Function
 $ I\left(\boldsymbol{k}_{T},\,\boldsymbol{q}_{T}\right) $   is given
 in Ref.\cite{LESI}. In this equation the expression in $\{\dots\}$ does
 not depend on $Q_T$ and stems from the short distances. In the same way
 as these short distance contributions cancel in the expression for
 $\sigma_{\rm eff}$ of \eq{DI11}, they will not contribute to the
 value of  $\sigma_{\rm eff} $   for double $J/\Psi$ or $\Upsilon$ productions.

 Therefore,  one can see from \fig{psi2sh} the single inclusive production
 of $J/\Psi$ is proportional to
 \bea \label{PSI1}
 \frac{d \sigma}{d y_1 } &\,\propto&\,g(Q_T=0) \int \frac{d^2
 \,Q_T}{4 \pi^2} N\Lb Q_T\Rb\, =\,\Lb \alpha^2 g_1 (Q_T=0) +
 \beta^2 g_2(Q_T=0)\Rb \int \frac{d^2 \,Q_T}{4 \pi^2} \Lb \alpha^2 \,g^2_1 (Q_T=0) \,+\,\beta^2 g^2_2(Q_T)\Rb\,\nn\\
 &=&\,\frac{1}{4 \pi}\Lb \alpha^2  g_1  + \beta^2 g_2\Rb\Lb\alpha^2 \frac{g^2_1}{B_1} \,+\,\beta^2\frac{g^2_2}{B_2}\Rb\,\,=\,\,\frac{1}{4 \pi}2 \Lb \alpha^2  g_1  + \beta^2 g_2\Rb\Lb\alpha^2 \frac{g^2_1}{B_1} \Rb \eea 
\eq{PSI1} is the generalization  to two channel model  of
 \eq{FD5}, which is written for the eikonal approach in which
 $N\Lb Q_T\Rb = g^2\Lb Q_T\Rb$.

 ~
 
\begin{figure}[ht]
    \centering
  \leavevmode
      \includegraphics[width=14cm]{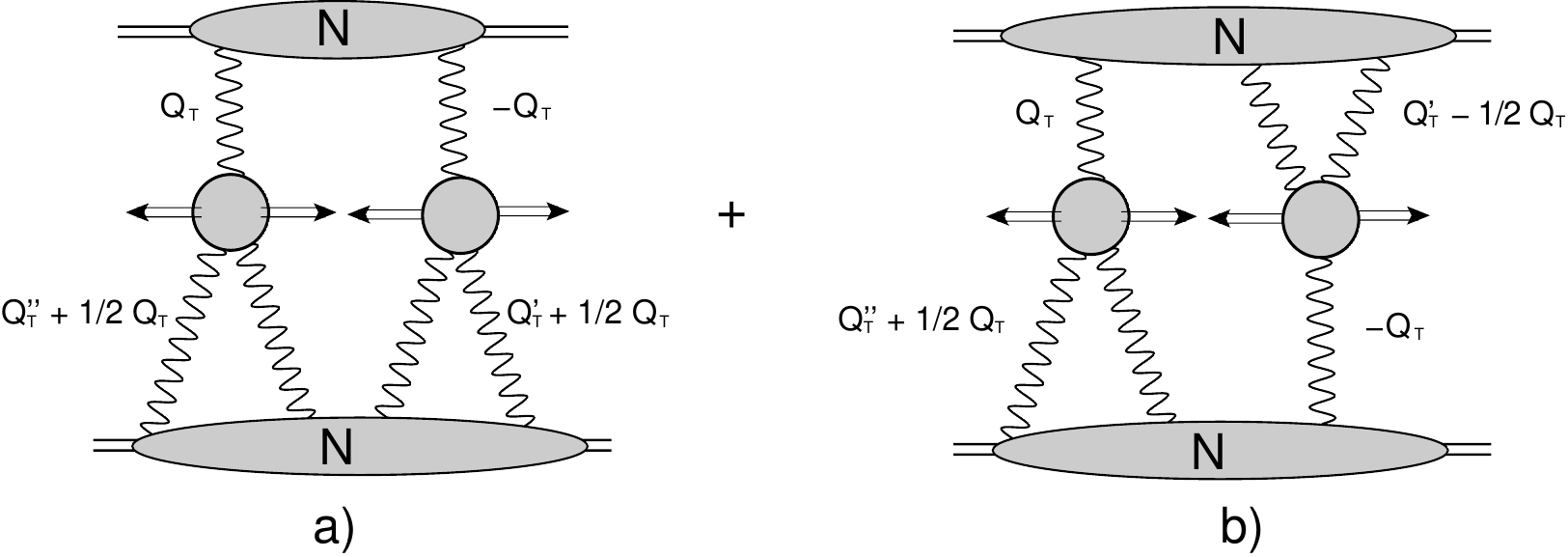}  
    \caption{Mueller diagrams\cite{MUDI} for the double
 inclusive cross section of pair $J/\Psi$ production.The wavy lines
 describe the BFKL Pomerons. The vertex V is defined in \eq{FD51}}
\label{psi2}
  \end{figure}

\begin{boldmath}
\subsubsection{Double inclusive productions   $J/\Psi$  and $\Upsilon$}
\end{boldmath}
The Mueller diagrams for the double inclusive cross section of
 $J/\Psi$ pair production are shown in \fig{psi2}. From this
 figure we see that the double inclusive cross section can be
 estimated using the amplitudes of the interaction with three
 and four Pomerons, which are shown in \fig{n}-b and \fig{n}-c.  Using these
 amplitudes, we obtain

\bea \label{PSI2}
&& \frac{d^2 \sigma}{d y_1\,d y_2 }\,\,=\,\,
\int \frac{d^2 Q_T}{4 \pi^2}\frac{d^2 Q'_T}{4 \pi^2}\frac{d^2 Q''_T}{4 \pi^2}
\Bigg(\Lb \alpha^2 g^2_1\Lb Q_T\Rb  + \beta^2 g^2_2\Lb Q_T\Rb\Rb\,\nn\\
&&\times\Bigg( \alpha^2 g_1\Lb \vec{Q}'_T + \h \vec{Q}_T\Rb g_1\Lb- \vec{Q}'_T + \h \vec{Q}_T\Rb g_1\Lb \vec{Q}''_T + \h \vec{Q}_T\Rb g_1\Lb - \vec{Q}''_T + \h \vec{Q}_T\Rb \nn\\
&&+ \beta^2 g_2\Lb \vec{Q}'_T + \h \vec{Q}_T\Rb g_2\Lb- \vec{Q}'_T + \h \vec{Q}_T\Rb g_2\Lb \vec{Q}''_T + \h \vec{Q}_T\Rb g_2\Lb - \vec{Q}''_T + \h \vec{Q}_T\Rb \Bigg)\nn\\
&& \,+\,\Bigg( \alpha^2 g_1\Lb \vec{Q}'_T + \h \vec{Q}_T\Rb g_1\Lb- \vec{Q}'_T + \h \vec{Q}_T\Rb g_1\Lb Q_T\Rb
+ \beta^2 g_2\Lb \vec{Q}'_T + \h \vec{Q}_T\Rb g_2\Lb- \vec{Q}'_T + \h \vec{Q}_T\Rb g_2\Lb Q_T \Rb \Bigg) \nn\\
&&\times \Bigg( \alpha^2 g_1\Lb \vec{Q}''_T + \h \vec{Q}_T\Rb g_1\Lb- \vec{Q}''_T + \h \vec{Q}_T\Rb g_1\Lb Q_T\Rb
+ \beta^2 g_2\Lb \vec{Q}''_T + \h \vec{Q}_T\Rb g_2\Lb- \vec{Q}''_T + \h \vec{Q}_T\Rb g_2\Lb Q_T \Rb \Bigg)\Bigg)\nn\\
&& \times\Bigg\{ \int d^{2}k_{T}d^{2}p_{T}\,\,G_{I\!\!P}^{\rm BFKL}\left(Y-y,\,p_{T},\,0\right)\,
\,\,V\Lb k_T,q_T\Rb\,G_{{I\!\!P}}^{{\rm BFKL}}\left(y;\,\boldsymbol{k}_{T}+\frac{1}{2}\boldsymbol{q}_{T},\,0\right)\,G_{{I\!\!P}}^{{\rm BFKL}}\left(y,-\boldsymbol{k}_{T}+\frac{1}{2}\boldsymbol{q}_{T},\,0\right)\Bigg\}^2\eea
 All momenta are shown in \fig{psi2}. In \eq{PSI2}  the factors in the curly brackets  depend only on short distances, which do not contribute in the estimates of $\sigma_{\rm eff}$.
Using  \eq{PSI1} and \eq{PSI2}
 we can calculate the values of $\sigma_{\rm eff}$ using \eq{XSEFF}:
 \beq \label{SIGMAF}
 \frac{1}{\sigma_{\rm eff}}\,\,=\,\, \frac{\frac{d^2 \sigma}{d y_1\,d y_2 }}{ \frac{d \sigma}{d y_1 }\,  \frac{d \sigma}{d y_2 } } 
 \eeq
 In this ratio all contributions from short distances cancels in the
 same way as in \eq{DI11} leading  to 
  the value of $\sigma_{\rm eff}
 $ that depends only on the contributions of the long distances in our
 processes.
 
 In \fig{xs4j}-b we plot the results of these estimates. Comparing
 \fig{xs4j}-a and \fig{xs4j}-b one can see that the values of
 $\sigma_{\rm eff}$ for $J/\Psi$ pair production, turns out to
 be smaller than 5 mb, which  is in accord with the experimental
 data of \fig{sigeff}.

 ~


\subsection{Conclusions}


In this letter we demonstrated  that the two parton showers mechanism
 leads to much smaller values of $\sigma_{\rm eff} \leq 5\,mb  $ for
the
 double parton interaction. This result is in qualitative agreement
 with the data, and we consider  it as  support for the idea that
 the production of two parton showers is responsible for  $J/\Psi$
  inclusive cross section.
The second result of this letter is the claim, that a simple two
 channel model with restrictions that stem from the results of the
 experiments on diffractive production in DIS,
is able to describe the size of the double parton interaction at high
 energies leading to the values of  $\sigma_{\rm eff}   \leq 20\, mb$ 
for
 4 jet production,  in accord with the experimental data. 

 As has been mentioned above, the value of  $\sigma_{\rm eff}$   is
 determined by the contribution of the long distances to the cross
 sections of   hard processes. These contributions are doomed to be
 modeled at present time using the phenomenological input, since out
 knowledge of the non-perturbative QCD is   still very limited.

Our numerical estimates are based on \eq{RES} and on the numerical
 values for the slopes $B_{el}$ and $B_{in}$.  We checked that the
 smallness of $\sigma_{\rm eff}$ for $J/\Psi$ pair production has   only
 a mild dependence on the value of $B_{in}$, which was measured with
 large errors. The two channel model has been used for describing the
 soft high energy data\cite{GLP}, It should be noted, that 
  for the BFKL Pomeron,  whose 
 intercept is larger than 1, the integral over the  produced mass
 in diffraction  is convergent, and
 the Good-Walker mechanism\cite{GW} is able to describe the
 diffractive  production  of both small and large masses\cite{GUGU}.
 Therefore, we believe  it reasonable to   use this model to obtain 
the first estimates
 of the value of $\sigma_{\rm eff}$.  It should be stressed that in
 the two channel model, that we used,  the parameter $\alpha$ 
determines 
the
 structure of the wave function of the hadron, and does not depend on the
 processes under consideration.

  {\it Acknowledgements.} \\
   We thank  all participants of Low-x 2019 WS  and especially L. Motyka for
 encouraging discussions.
 This research was supported  by 
 CONICYT PIA/BASAL FB0821(Chile)  and  Fondecyt (Chile) grant 1180118 .

\end{document}